\documentclass[twocolumn]{IEEEtran}
\IEEEoverridecommandlockouts
\usepackage{cite}
\usepackage{amsmath,amssymb,amsfonts}
\usepackage{algorithmic}
\usepackage{graphicx}
\usepackage{textcomp}
\usepackage{xcolor}
\usepackage[ruled,vlined]{algorithm2e}
\def\BibTeX{{\rm B\kern-.05em{\sc i\kern-.025em b}\kern-.08em
    T\kern-.1667em\lower.7ex\hbox{E}\kern-.125emX}}
\begin{document}

\title{I-nteract: A cyber-physical system for real-time interaction with physical and virtual objects using mixed reality technologies for additive manufacturing
\thanks{This publication has emanated from research supported in part by a research grant from Science Foundation Ireland (SFI) under grant number 16/RC/3872 and is co-funded under the European Regional Development Fund and by I-Form industry partners.}
}

\author{\IEEEauthorblockN{Ammar Malik\IEEEauthorrefmark{1}, Hugo Lhachemi\IEEEauthorrefmark{1}, Robert Shorten\IEEEauthorrefmark{1}\IEEEauthorrefmark{2}}\\

\IEEEauthorblockA{\IEEEauthorrefmark{1}School of Electrical and Electronic Engineering, University College Dublin, Dublin, Ireland \\
ammar.malik@ucdconnect.ie, hugo.lhachemi@ucd.ie\\
\IEEEauthorrefmark{2}Dyson School of Design Engineering, Imperial College London, London, United Kingdom\\
r.shorten@imperial.ac.uk}
}
\graphicspath{{./Figures/}}
\maketitle

\begin{abstract}
This paper presents I-nteract, a cyber-physical system that enables real-time interaction with real and virtual objects in a mixed augmented reality environment to design 3D models for additive manufacturing. The system has been developed using mixed reality technologies such as HoloLens, for augmenting visual feedback, and haptic gloves, for augmenting haptic force feedback. The efficacy of the system has been demonstrated by generating 3D model using a novel scanning method to 3D print a customized orthopedic cast for human arm, by estimating spring rates of compression springs, and by simulating interaction with a virtual spring using hand. 
\end{abstract}

\begin{IEEEkeywords}
Additive Manufacturing, Cyber-Physical System, Haptics, Human-Computer Interaction, Mixed Reality.
\end{IEEEkeywords}

\section{Introduction}
\label{sec:introduction}
During the last three centuries, changes in manufacturing introduced by the three first industrial revolutions have acted as a catalyst for profound societal change. Modern societies are now entering into a fourth industrial revolution, called Industry 4.0, driven by ubiquitous connectivity and which is characterized by the emergence of smart factories and smart services providers~\cite{lee2015cyber}, and by developing smart products and smart services~\cite{lee2014service}. As a part of this global trend, the last decade has seen the emergence of additive manufacturing (AM) as a disruptive technology poised to deeply transform manufacturing~\cite{cotteleer20143d,stock2016opportunities}. AM, commonly known as 3D printing, refers to the various processes of adding together materials, based on a 3D model file, for producing three-dimensional objects. Originally used for fast prototyping~\cite{RN41,RN82,RN83}, AM has seen rapid growth during the last decade due to the advancement in processes and tools, as well as a number of appealing features such as the obviation of the need for dedicated tooling and the capability to manufacture bespoke products with specific shapes or advanced features that are not manufacturable with traditional manufacturing methods. Moreover, the emergence of low-cost 3D printers that are affordable for the general public is a game changer as it narrows the gap between the consumer, the designer, and the production. In the global context of climate change, such a democratization of accessible production units might also play a key role in the emergence of a more sustainable economy that includes repair, upgrade, and refurbish as a service. 

Despite tremendous promises, AM still faces many challenges~\cite{oropallo2016ten}. Much research work remains to be done in order that AM matches the standards of conventional manufacturing and reaches its full potential. In particular, from a control design perspective, the design of cyber-physical feedback control systems and corresponding support tools has become a key enabler toward its both popularisation for the general public and widespread adoption by the industry~\cite{lhachemi2019augmented}. The objective in this context are threefold: make 3D printing machines work better (low-level control), make the human and the 3D printer work efficiently together (human-in-the-loop), and make the 3D printers work efficiently together over a network (machine-machine orchestration). 

The objective of this paper is to contribute to the improvement of the traditional 3D printing workflow by enabling a better interaction between the human and the machine. The traditional 3D printing workflow goes as follows. Starting with the specifications, a 3D model of the part is designed via a computer aided design (CAD) tool. The resulting 3D model is printed and then tested. Depending on the results of the tests, the design loop is iterated in order to reach the desired level of functionality of the printed part. Such a design loop is sub-optimal. First, the interaction of the human with the 3D model is limited by a computer set (keyboard, mouse, 2D screen), which is essentially counter intuitive for positioning and modelling within a three-dimensional environment~\cite{RN87,RN53}. Second, the printing of the part is costly (time and material) and the cost increases with the number of iterations of the design loop. Finally, most of the test of the printed part is postponed to the end of the printing process. As a result, designing 3D printed products is presently inefficient and difficult. This is a major obstacle to widespread adoption and home use~\cite{RN196}. 

\begin{figure*}
    \centering
    \includegraphics[]{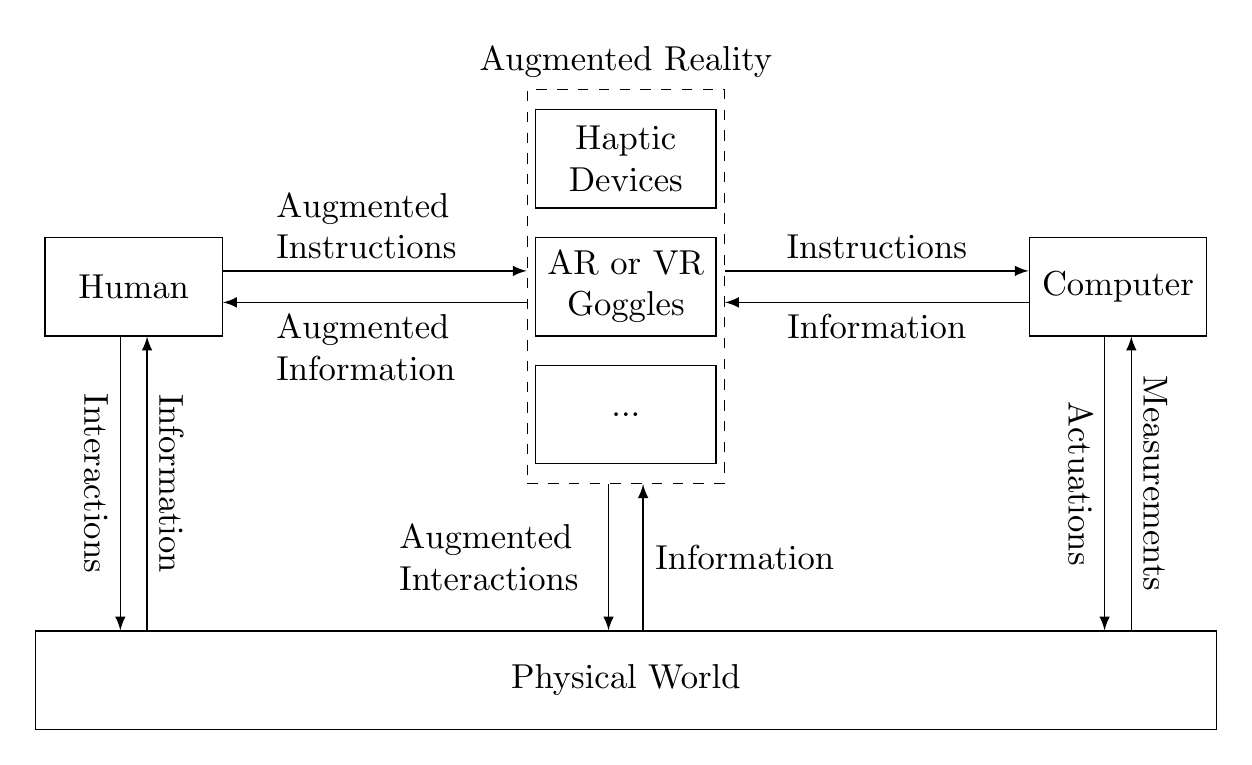}
    \caption{Feedback based on augmented reality~\cite{lhachemi2019augmented}}
    \label{fig:Fb based on augmented reality}
\end{figure*}

The use of augmented reality (AR) tools and haptic feedback has been identified as a promising approach for helping consumers to use 3D printing technologies in a better way than is currently possible~\cite{lhachemi2019augmented}. As shown in Fig.~\ref{fig:Fb based on augmented reality}, AR technologies allow to reduce the gap between the digital nature of the 3D model, existing only in the virtual world, and the tangible nature of the final 3D printed product. They can be used as a key enabler to facilitate interactions for assembling, creating, interacting, modifying, positioning, and shaping 3D models within a three-dimensional environment. Moreover, haptic technologies can be used in order to enable better interaction with virtual objects. Indeed, contrary to traditional human-machine interface that take advantage of visual and auditory senses, haptic devices are used to generate mechanical feedback stimulating the sense of touch. 

This paper presents I-nteract. The system enables the user to interact with virtual objects for AM. The developed cyber physical system (CPS) combines the AR technology Hololens\footnote{https://docs.microsoft.com/en-us/hololens/hololens1-hardware}, commercialised by Microsoft, and the haptic Glove Dexmo\footnote{https://www.dextarobotics.com/en-us}, commercialised by Dexta Robotics based on the early prototype~\cite{gu2016dexmo}. I-nteract allows the user to interact with virtual artifacts taking the form of holograms while receiving force feedback for an enhanced interaction. These interactions include translations, rotations, perception of elasticity, etc. The system also allows the interaction with physical objects for 3D model generation and elasticity estimation. The model generation tool is based on a hand-based novel scanning method that enables the user to generate 3D models in real-time to design customized products for personal fabrication. The efficiency of I-nteract is demonstrated by designing and printing a customized orthopedic cast for a human arm.

The remainder of this paper is organized as follows. Related works are presented in Section~\ref{sec:Related works}. After a general presentation of the system in Section~\ref{sec:System overview}, the different interactions allowed by I-nteract are descibed in Section~\ref{sec:Interaction}. Case studies are reported in Section~\ref{sec:Results}. Finally, concluding remarks are formulated in Section~\ref{sec:Conclusion}

\section{Related works}
\label{sec:Related works}

The use of AR technologies in the context of 3D modeling and additive manufacturing has emerged as a very active research direction. A complete review regarding the use of such technologies to improve the user's interaction capabilities is reported in~\cite{lhachemi2019augmented}. The present section focuses on the specific contributions that allow gesture-based interactions with artifacts, interactions between virtual and real objects, as well as haptic feedback.

Gesture-based interactions~\cite{RN20,RN21,RN36} are among the most promising approaches for improving human-machine interactions in the context of 3D modelling. Requiring real-time hand tracking and system processing to recognize human gestures, this type of system has emerged multiple times in the recent literature. FingARtips~\cite{RN10} is a fingertip-based AR interface combining visual markers and vibrotactile actuators. Tangible~3D~\cite{RN17} is an immersive {3D} modeling system, 
using cameras and projectors to create and interact with {3D} models. User's gestures have been used in~\cite{RN14} to propose Data Miming, a system that allows the user to describe physical objects with gestures. 

AR interfaces have proved to be more intuitive for manipulating {3D} models. However, gesture-based only modelling approaches suffer from the fact that existing head-mounted displays are unable to provide a sufficient level of precision due to mid-air gesture-based inputs. The use of real world objects in the virtual environment as a reference for physical guidance has emerged has a promising approach~\cite{RN32}. MirageTable~\cite{RN9} combines a depth camera, a curved screen, and a stereoscopic projector to merge real and virtual worlds. Another example of {3D} modeling system has been developed in~\cite{RN46}. This system uses an aerial imaging plate and a leap motion sensor to manipulate (move, scale and rotate) the virtual object using gestures to fit the physical object~\cite{RN46}. 

Interaction with virtual objects can be counter-intuitive when relying solely on visual feedback. Indeed, contrary to real world interactions, interactions with virtual objects do not generate any mechanical feedback. In this context, the use of tangible tools (including additional hardware) for the creation or modification of virtual models has been identified as a possible path to make interactions with the virtual world more intuitive.  Surface drawing~\cite{RN22} proposes, in combination with a hand tracker, the use of physical tools (tongs, erasers, magnetic tools) for the shape refinement of {3D} models. Twister~\cite{RN18} introduces a hand-based manipulating tool that allows the user to tilt, twist, and bend {3D} shapes. A wrist-worn sensor, named Digits, was developed in~\cite{RN22}. This system estimates the {3D} pose of the user's hand, enabling natural hand manipulations in the virtual environment. Taking advantage of interactive situated AR systems like HoloDesk~\cite{RN13} and Holo Tabletop~\cite{RN45}, MixFab~\cite{RN1} is an immersive augmented reality environment that allows the user to sketch and extrude the virtual artifact using hands. This system combines a depth camera to recognize hand gestures and a motorized turntable for {3D} scanning of the physical object. This enables the user to use a digital model of the scanned physical object as a size or shape reference to integrate the physical object in the design process.
 
Beyond the use of tangible objects and tools for improving interactions with the virtual environment, the use of haptic devices has emerged as a very promising research direction~\cite{RN68}. Haptic interfaces are devices that generate mechanical signals to stimulate kinesthetic~\cite{RN2,RN61,RN63,RN73} and/or tactile senses~\cite{RN76,RN69,RN72} of the human. In the context of this work, we are mainly interested in haptic devices providing a kinesthetic feedback to the user as this type of feedback can be used to perceive size, weight, and position of the object in the virtual environment. For instance, such kinesthetic feedback can be used to prevent the user hands from penetrating through the virtual object and hence, provide a more realistic experience. Among the available technological solutions, exoskeletons have demonstrated there efficiency as haptic interfaces, allowing the generation of complex arrangements of force feedback~\cite{bergamasco2007exoskeletons,zhou2014rml}. 

Haptic feedback technologies offer promising solutions for the improvement of AR experiences in additive manufacturing. For instance, the work reported in~\cite{RN50} uses a haptic device for navigating in CAD environments. Specifically, the user can manipulate the camera in the 3D environment by means of a tangible tool providing force feedback when a virtual obstacle is encountered. Another example is reported in~\cite{RN70}, where the use of different feedback methods that include pressure-based tactile and vibrotactile feedbacks have been developed for improving human-machine interactions. With specific reference to this work, one solution in this direction is the haptic Glove Dexmo commercialised by Dexta Robotics. This commercialised version is based on the early prototype reported in~\cite{gu2016dexmo}. While proprietary basic low-level control laws (inner-loop) have been accomplished by Dexta, research activities are currently on-going to provide a more complete command strategy (outer-loop), either in term of force or position feedback~\cite{friston2019position}.

\section{System Overview}
\label{sec:System overview}

\begin{figure*}
    \centering
    \includegraphics[width=7in]{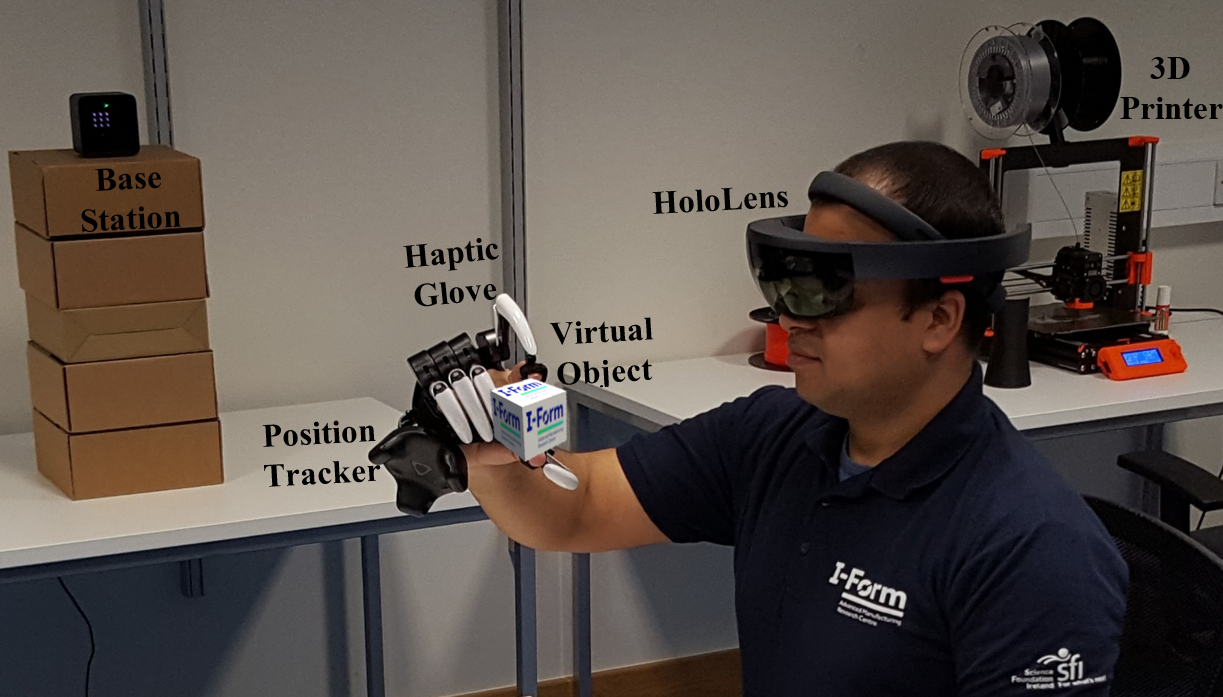}
    \caption{I-nteract}
    \label{fig:sys}
\end{figure*}

We now present I-nteract. I-nteract is the system that, for the first time, to the best of our knowledge, uses MR and haptic feedback to create MR environment in which (deformable) virtual and physical objects can interact with each other while providing an integrated visual and haptic (visio-haptic) experience to the user for AM. The setup of the I-nteract is shown in Fig \ref{fig:sys}. It includes mixed reality (MR) glasses (HoloLens) for visual feedback, haptic gloves (Dexmo) for force feedback, and VIVE\footnote{https://www.vive.com/ca/vive-tracker/} hardware for global position tracking of a glove. Dexmo is a haptic exoskeleton glove that provides kinesthetic feedback to interact with the virtual objects. The kinematic design of the glove is similar to the human hand kinematics model developed in \cite{meng2011development}. The kinematic design consists of the rigid links and joints. Each glove's digit has two rigid links (primary and secondary bar) and three joints (MCP, PIP, and DIP joints). The primary bar in each digit connects MCP (metacarpophalangeal) and PIP (proximal interphalangeal) joints whereas the secondary bar connects PIP and DIP (distal interphalangeal) joints. The MCP joint of the thumb has three DOF (rotate, split and bend) whereas the MCP joint of each finger has two DOF (split and bend). The PIP and DIP joints of each digit has only one degree of freedom (DOF). Therefore, the glove has 21 DOF \cite{friston2019position}. The actuators (motors) that provide force feedback and the position sensors that capture the fingers motion are installed only at the MCP joints of each digit. Therefore, the glove is under-actuated (5 DOF) and under-instrumented (11 DOF). The position sensor readings, ranging from 0 to 1, are normalized values between the extremes of the user's finger flexion and extension. Using position sensors Dexmo capture the fingers motion and updates the virtual model of the hand accordingly. When this virtual hand comes in contact with a virtual object, Dexmo's API calculates direction and amplitude of the forces and sends commands to the actuators to provide the force feedback to the five fingertips. The force feedback can also be manually controlled using two parameters, namely: \textit{stiffness} and \textit{offset}. The stiffness parameter, ranging from 0 to 5, determines the strength of the force feedback. The offset parameter, ranging from 0 to 1, determines the position at which force feedback is applied.  

\subsection{Virtual hand representation}
In Dexmo's API, the complete hand model is composed of three sub hand models, namely: \textit{Graphics Hand Model}, \textit{Trigger Collider Hand Model}, and \textit{Collision Collider Hand Model}. \textit{Graphics Hand Model} is the virtual hand model that contains all the meshes to display the 3D model of the virtual hand. \textit{Trigger Collider Hand Model} contains all the trigger colliders that are used to calculate force feedback based on the colliders attached to the virtual object with which the user is interacting. \textit{Collision Collider Hand Model} contains all the non-trigger colliders that are used as the kinematic rigid body when hand model is moving. These colliders are employed to simulate physical interactions using Unity's physics engine. For instance, push away other virtual objects which also have colliders attached to them.

\subsection{Global position tracking of the hand}
\begin{figure}
    \centering
    \includegraphics[width=3.5in]{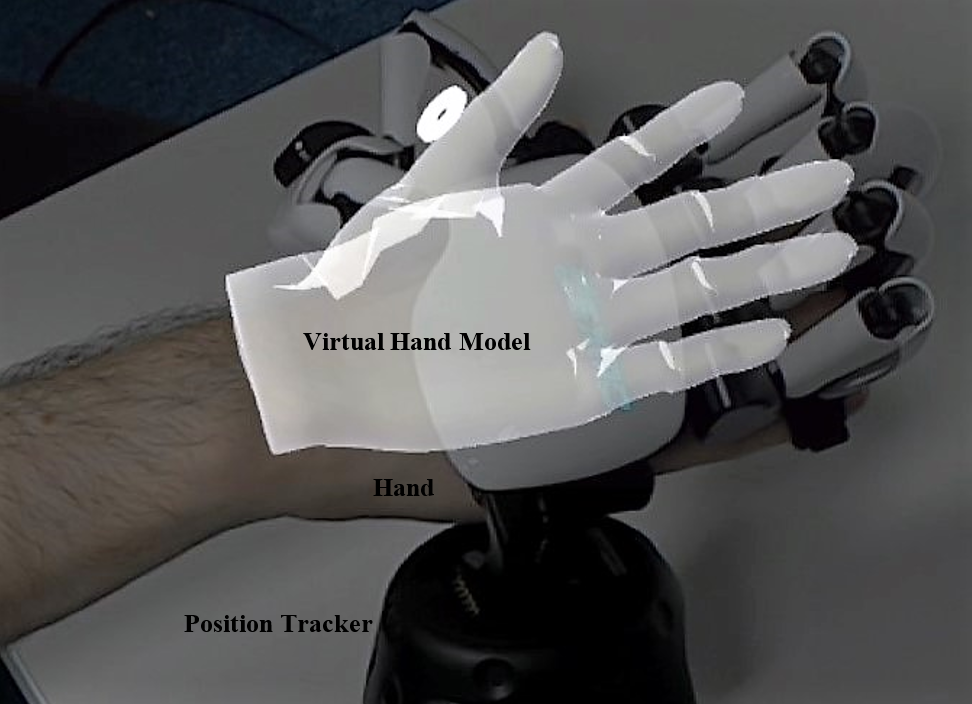}
    \caption{Placing hand on the virtual hand model for synchronization}
    \label{fig:sync}
\end{figure}

In addition to 21 DOF that represents finger motion of the glove's digits with respect to the palm provided by Dexmo API, we have added 6 DOF for global position tracking of hand. The global position tracking is required to synchronize the user's hand with the virtual hand model so that, when the user moves his/her hand in a physical work space to interact with virtual or physical objects, the virtual hand model moves with it. HoloLens uses physical position of the user at the starting point of the app as world origin and projects the virtual objects in world space accordingly. In order to track the position and orientation of the hand in the user workspace with respect to the world origin, we have used Vive hardware (base station and tracker). The global position tracking of the hand has been adapted from \cite{BoxAndBlocks}. The position tracker attached to the glove is tracking position with respect to the base station coordinates. As the base station orientation and position is different from the orientation and position of the world origin, it is necessary to synchronize the base station coordinates and the world coordinates. To synchronize the user's hand position and orientation (base station coordinates) with the virtual hand (world coordinates), the user places the hand on the virtual hand model as shown in Fig \ref{fig:sync} and uses the voice command \textit{sync}. As soon as the virtual hand model is synchronized with the user's hand the \textit{Graphics Hand Model} is deactivated.    

\section{Interaction}
\label{sec:Interaction}
Apart from inaccessibility to a large non-technical consumers, most contemporary CAD-based software demands a strong technical background. Even the availability of the open source 3D modelling software on the internet  does  not  solve  the inherent difficulty of human-machine interactions for manufacturing. The use of traditional mouse and keyboard to locate and place objects in a virtual 3D environment with accessibility through the 2D window of a computer screen is inherently difficult. Therefore, this traditional interface for interaction is counter-intuitive for assembling,  creating,  interacting,  modifying,  positioning, and shaping 3D models within a three-dimensional environment. In this context, there is a clear need for developing new interfaces that take advantage of mixed reality technologies for interacting with 3D models as part of the design process~\cite{lhachemi2019augmented}. I-nteract is an interface for real-time interaction with both real and virtual objects that enables kinematic and dynamic interaction for AM. 

In this paper, we consider two types of interactions, namely: kinematic and dynamic interactions. A kinematic interaction is a type of interaction in which the shape of the object (whether virtual or real) does not change when a force is applied by the user. Conversely, a dynamic interaction is a type of interaction in which the shape of the object does change when a force is applied by the user.

\subsection{Kinematic Interactions}

\subsubsection{Position and orientation manipulation of virtual objects}

\begin{figure}
    \centering
    \includegraphics[width=3.5in]{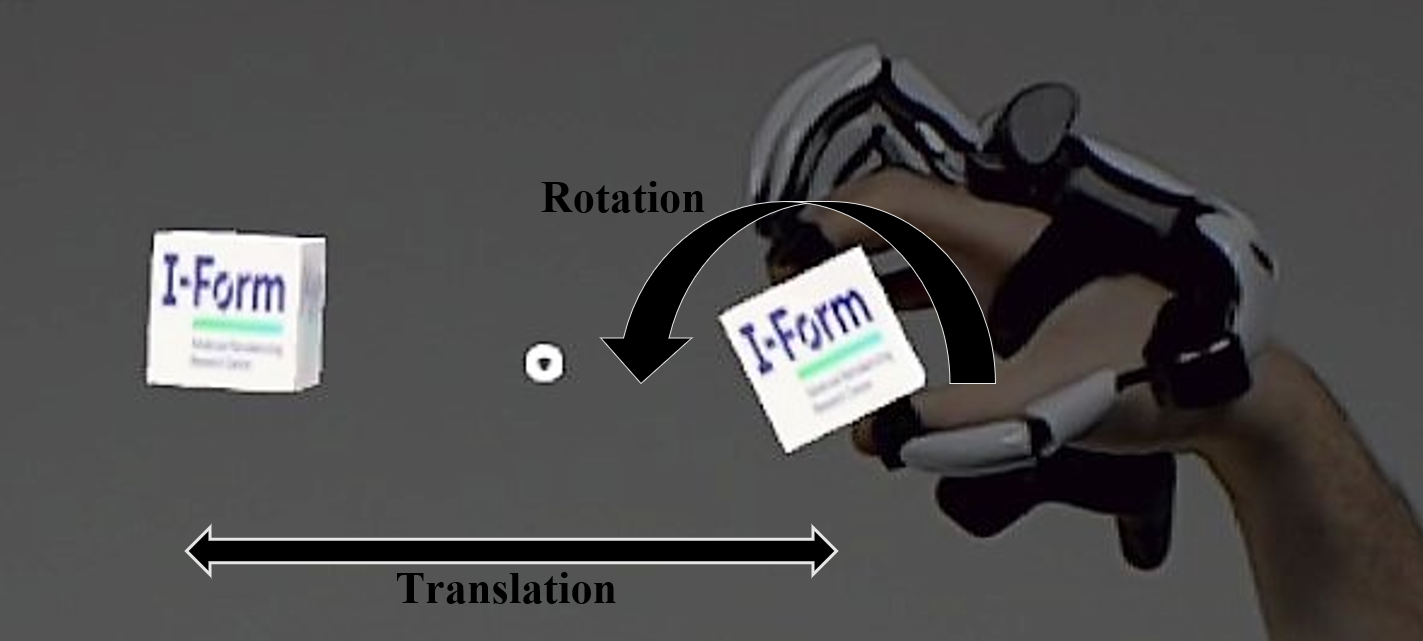}
    \caption{Moving and rotating a virtual object in 3D space}
    \label{fig:cube}
\end{figure}

 As illustrated by Fig.~\ref{fig:cube}, I-nteract allows the user to rotate, move and get force feedback according to the contours of the virtual object. This interaction with virtual objects can not only be used for design purposes to modify the 3D model, but can also be used for monitoring purposes during 3D printing. In~\cite{malik2019}, a 3D model reconstruction for the real-time monitoring of additive manufacturing processes is proposed. An AR interface was developed which enabled the user to interact with the reconstructed 3D model using gestures to view and detect potential defects, not only at the surface but also in the inner layers of the printed object during the printing process. I-nteract provides a more intuitive interface, in which the user can interact with the reconstructed 3D model by rotating and moving in 3D AR space using hands instead of gestures to monitor the printing process. 

\subsubsection{3D model generation of a physical object}
\begin{figure}
    \centering
    \includegraphics[width=3in]{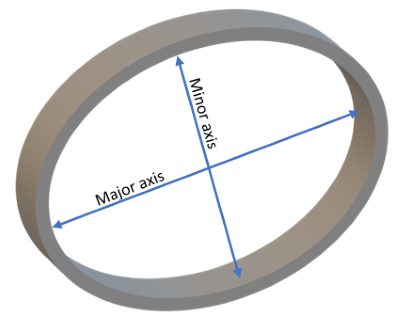}
    \caption{Hollow elliptic cylinder 3D model for shape estimation}
    \label{fig:cylinder}
\end{figure}

In this paper, we present a novel scanning method to generate a 3D model of a physical object by estimating its contours using a haptic glove. HoloLens is used for MR visual feedback and implementation of the voice commands. The position sensors at the MCP joints of all the glove's digits are used to estimate the hand pose as well as position of the fingertips. Therefore, the user can capture the contour of a physical object using the data from the position sensors (used for position tracking of each finger with respect to the glove palm) and the position tracker (used to track the position of glove in the user's 3D space). Traditional scanning techniques require either scanners or the physical objects to move around each other. Therefore, the scanning strategy developed in this paper is suitable for the scanning of the physical objects which have restrictive movements. An example of such a situation is a patient with fractured bone. We have implemented this strategy to 3D print a customized orthopedic cast for human forearm. As the human limbs present a cylindrical shape, we have used virtual hollow elliptical cylinders to estimate the contours of the limbs; see Fig.~\ref{fig:cylinder}. 

\begin{figure}
    \centering
    \includegraphics[width=3.5in]{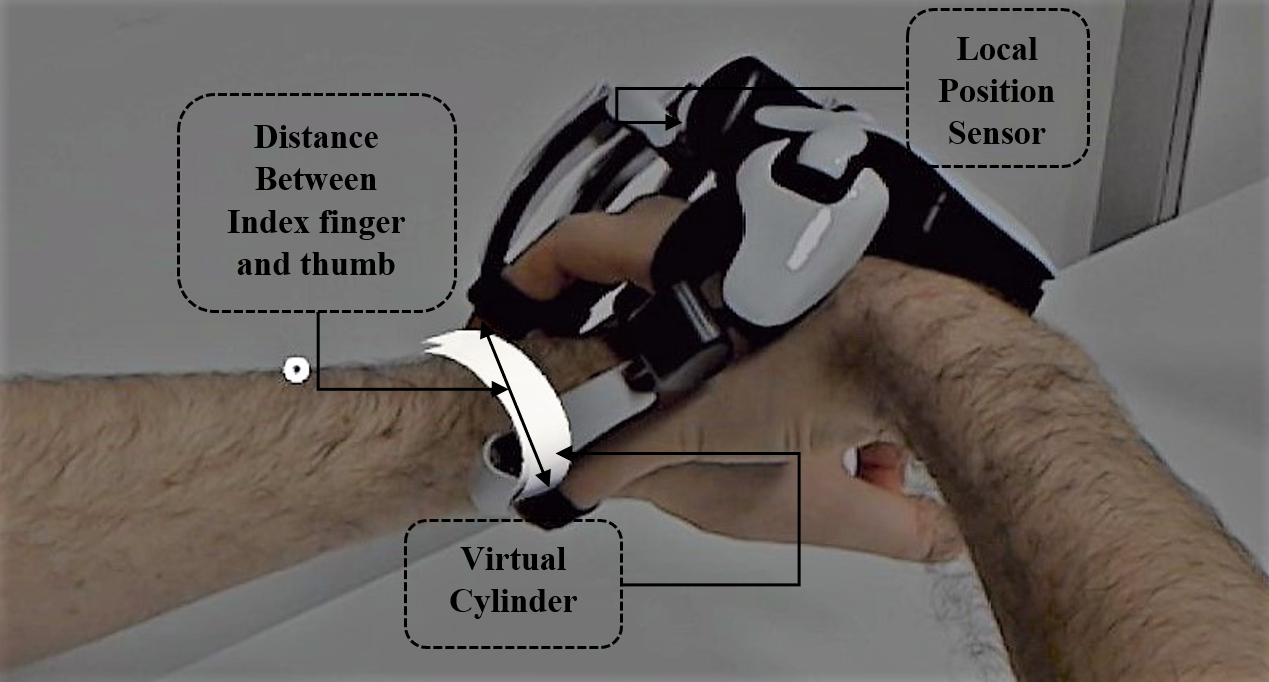}
    \caption{Fitting cylinder on the segment of forearm}
    \label{fig:cylinder_fit}
\end{figure}

The procedure implemented for generating a 3D model of the orthopedic cast is described below.
\begin{enumerate}
    \item Place the thumb and the index fingertip onto the segment of arm to fit a virtual cylinder horizontally as shown in Fig.~\ref{fig:cylinder_fit}.
    \item Say ''Vertical''.
    \item Place the thumb and the index fingertip onto the segment of arm to fit the cylinder vertically.
    \item Say ''Next''.
    \item Repeat to step 1 to 4 until fitting the cylinder onto the last segment.
\end{enumerate}

\begin{figure}
    \centering
    \includegraphics[width=3.5in]{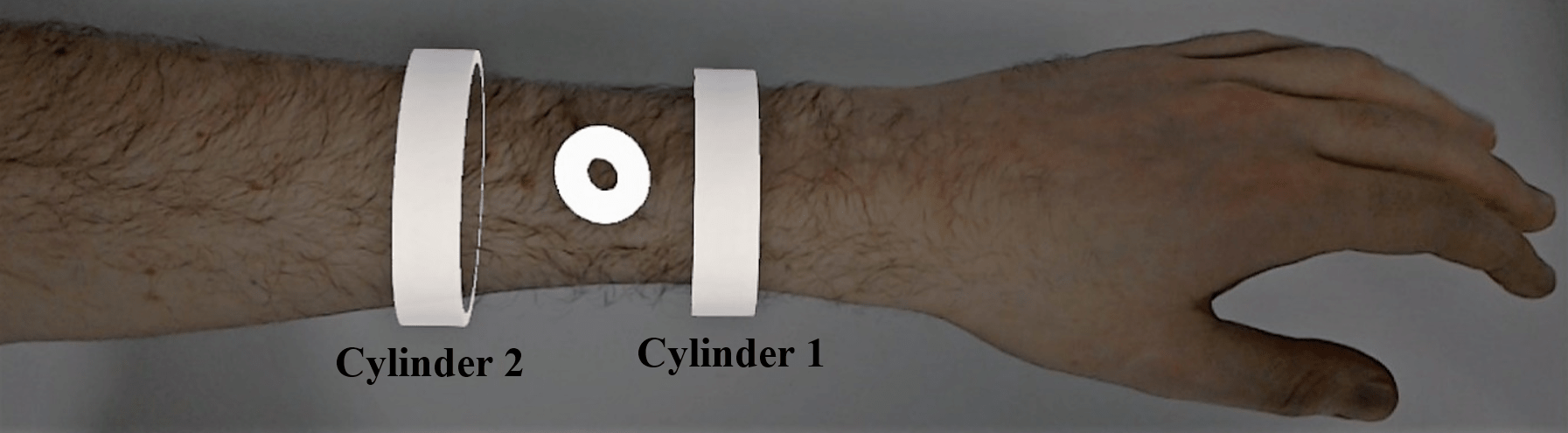}
    \caption{Fitting cylinders at two segments of forearm}
    \label{fig:cylinder1_2}

    \includegraphics[width=3.5in]{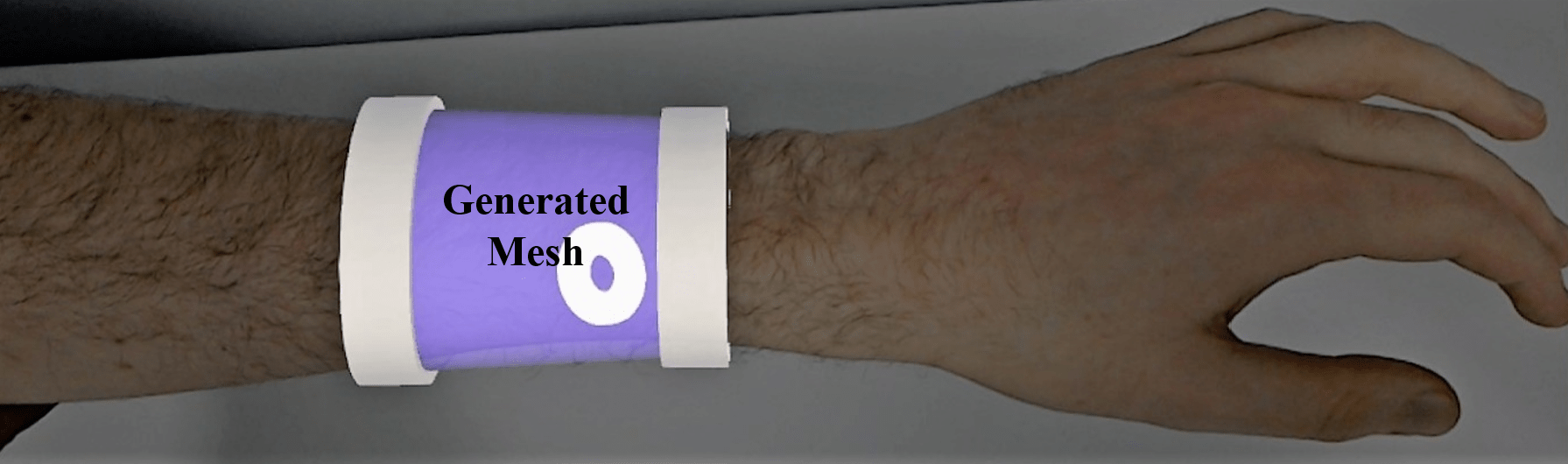}
    \caption{Mesh generated based on measurements at two segments}
    \label{fig:Mesh1}

    \includegraphics[width=3.5in]{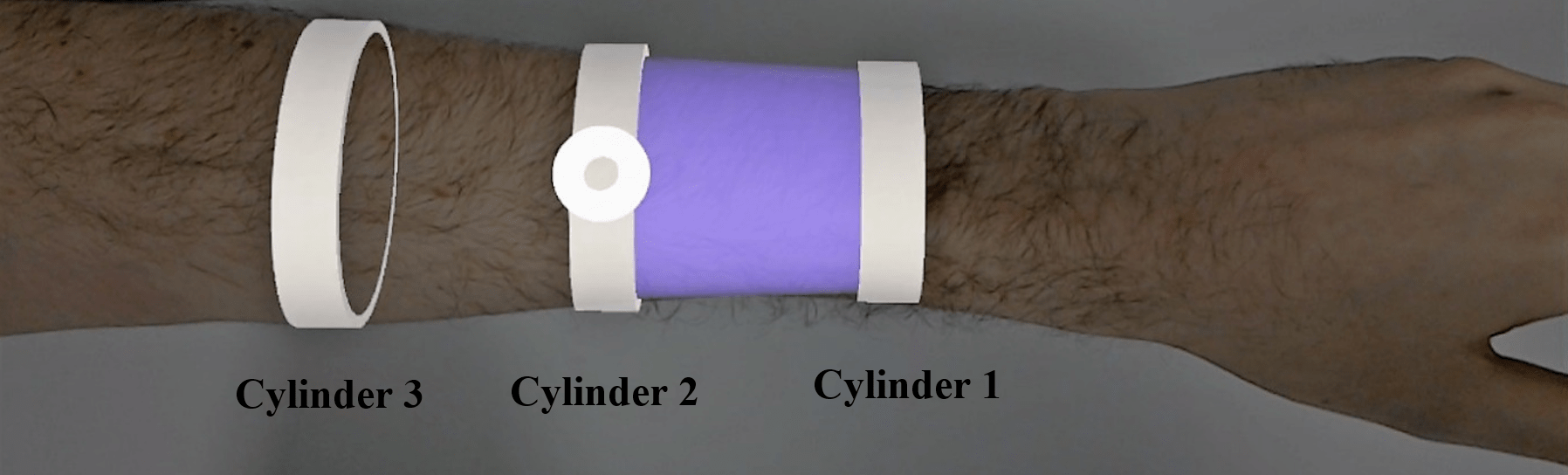}
    \caption{Fitting cylinder at third segment}
    \label{fig:cylinder3}

    \includegraphics[width=3.5in]{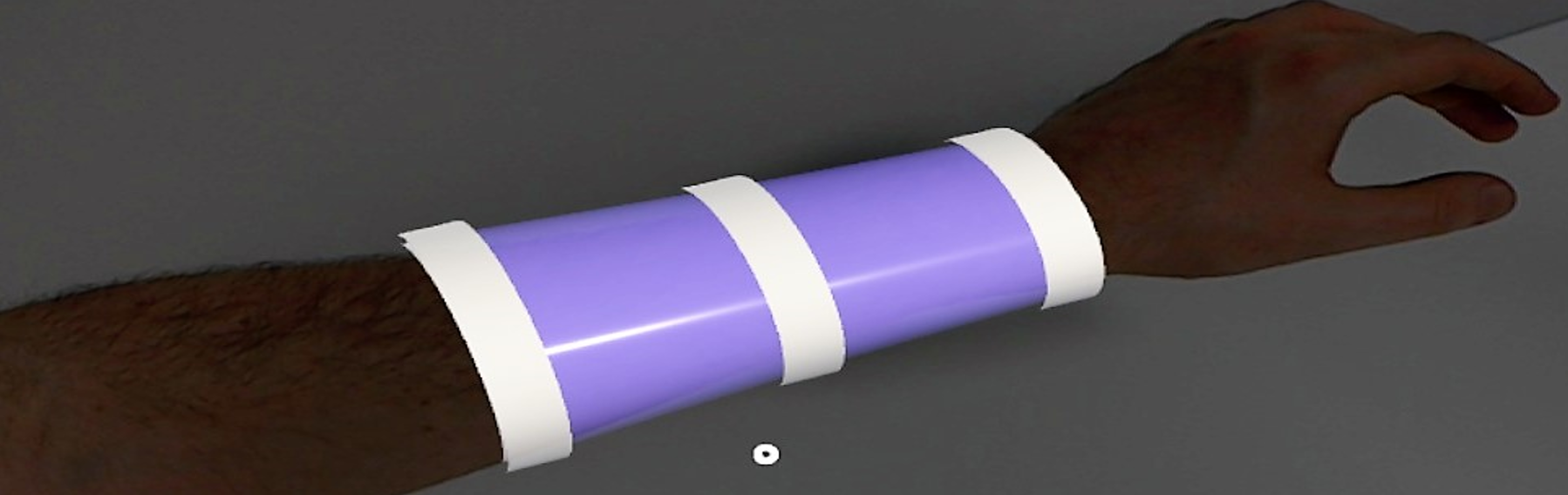}
    \caption{Mesh generated based on measurements at three segments}
    \label{fig:Mesh2}
\end{figure}

The user can fit as many cylinders as the user wants onto the different segments of the arm to capture its shape. We have generated the 3D model by fitting the cylinders onto the three segments of arm with a distance of 7.5~cm between the consecutive cylinders. Figure~\ref{fig:cylinder1_2} shows the first two cylinders fitted onto the arm. Figure~\ref{fig:Mesh1} shows the generated mesh between the two fitted cylinders and hence capturing the shape of the arm between the two segments. Figure~\ref{fig:cylinder3} shows the third cylinder fitted onto the third arm segment. Finally, Fig.~\ref{fig:Mesh2} shows the generated mesh between the second and the third fitted cylinders. 

The next paragraphs describe the three main parts of the implementation for the generation of the 3D model of the cast.

\begin{figure}
    \centering
    \includegraphics[width=3.5in]{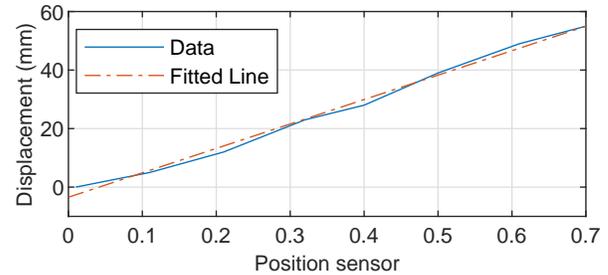}
    \caption{Position sensor reading and displacement mapping}
    \label{fig:linear_map}
\end{figure}

\textbf{Distance estimation using position sensor.}
We have used position sensor readings of the glove's index finger to estimate the distance between the thumb and the index finger tip. This enables the user to take measurements of the physical object at various segments by placing the index fingertip at one end and thumb at the other end of the segment. For a fixed finger pose (that is for fixed PIP and DIP joint angles), the relation between finger tip position and the position sensor can be linearly mapped using linear regression. We have measured the distance (between the thumb and index finger tip) using vernier caliper for each position sensor value (that is from 0 to 1 with step size of 0.01) and have used this data to find the line of best fit. This linear mapping is shown in Fig.~\ref{fig:linear_map}.

\textbf{Fitting virtual cylinders.}
The contour of the physical object is estimated by fitting virtual hollow elliptic cylinders (Fig.~\ref{fig:cylinder}) onto the segments of the object. The major and minor axes of the virtual cylinder are programmed to vary according to the distance between the thumb and the index finger tip in real time. As the user moves index finger relative to the thumb, the major or the minor axis changes according to the index finger position sensor reading. The user can visualize the change in the shape of the cylinder via HoloLens in real time and hence can fit the cylinder according to the height and the width of the segment. In Fig.~\ref{fig:cylinder_fit} the user is fitting the cylinder horizontally. Each cylinder is fitted onto the segments by placing the index finger and the thumb on the segment either horizontally or vertically. The voice command control has been implemented using HoloLens to enable the user to switch between changing the major axis and the minor axis to fit the cylinder both horizontally and vertically on the segment of the arm.

\begin{figure}
    \centering
    \includegraphics[width=3.5in]{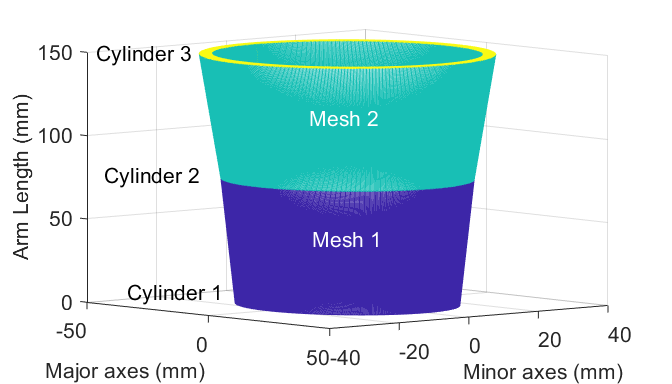}
    \caption{Mesh interpolation to estimate contour}
    \label{fig:Mesh}
\end{figure}

\textbf{Mesh Generation.}
The mesh is generated between the two cylinders as soon as the cylinders are fitted on the arm. Figure~\ref{fig:Mesh} shows the generated mesh based on the fitted cylinders onto the arm. Algorithm~\ref{alg:Algo} has been used to compute faces of the triangles based on the vertices of the consecutive fitted cylinders. The faces and vertices for each consecutive cylinders are then written to an OBJ file to generate (mesh) the 3D model of the cast.

\begin{algorithm}[t!]
\SetAlgoLined
v1 = [x1, y1, p1]\; \tcp{v1 = vertices of cylinder 1}
v2 = [x2, y2, p2]\; \tcp{v2 = vertices of cylinder 2}
\tcp{dimension of v1 or v2 = number of vertices $\times$  3}
\tcp{p1 = position of cylinder 1 on arm}
\tcp{p2 = position of cylinder 2 on arm}
 \For{i = 1 to n-1}{
 \tcp{n = number of vertices in one cylinder}
  f1=[v2(i),v1(i),v1(i+1)]\;
  f2=[v1(i+1),v2(i+1),v2(i)]\;
  face=[face;f1;f2]\;
 }
 \caption{To compute triangles' faces for mesh generation}
 \label{alg:Algo}
\end{algorithm}

\subsection{Dynamic Interaction}
Due to the development of AR technologies, the dynamic simulation of deformable objects in real-time for interactive MR environments is an active area of research~\cite{natsupakpong2010determination}. Springs constitute the basic building blocks to estimate and simulate the elasticity (or stiffness) of a deformable object~\cite{chang2006deformable,sung2017simulation,vassilev2002mass}. In this section, we present a method, first to estimate the spring rate (stiffness) of a real compression spring, and second to simulate the interaction with a virtual compression spring, using I-nteract.

\subsubsection{Elasticity estimation of real springs}

\begin{figure}
    \centering
    \includegraphics[width=3.5in]{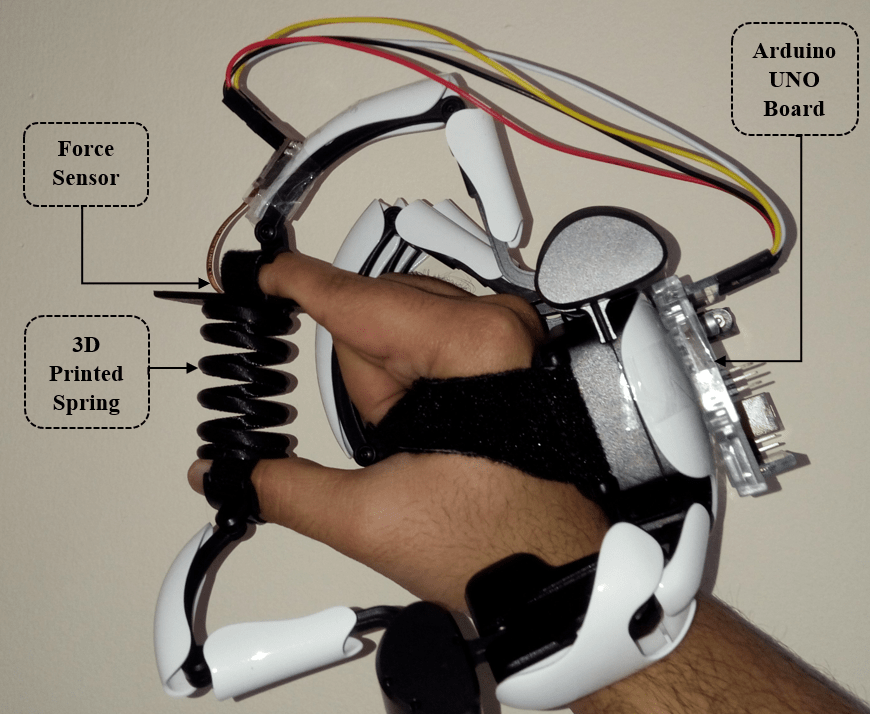}
    \caption{User holding a 3D printed spring to estimate spring rate}
    \label{fig:g_spring}
\end{figure}

Elasticity is the property of an object that causes it to resist a deforming influence and be restored to its original shape when the force causing the deformation is removed. A compression spring is an elastic object which exerts, when compressed, a restoring force to get back to its original length. The restoring force is proportional to the displacement in spring's length caused by compression and is described by Hooke's Law: 
\begin{equation}\label{eq:HLaw}
  F = k x  ,
\end{equation}
where $k$ is the spring constant, $F$ is the restoring force of the spring and $x$ is the relative spring displacement with respect to its free length\footnote{Note that we are only concerned by replicating the restoring force of the spring. In particular, we do not intend here to replicate the dynamical behavior of a mass-spring system (which could be done by adding accelerometers to the current system). Considering the targeted application, namely 3D modelling for AM, such a simplification assumption is reasonable because of the relatively low speeds and accelerations that can be achieved by a human fingertip.}. Using I-nteract, the user can estimate the spring constant of a real spring. To do so, the user compresses the spring multiple times using the index finger and thumb as shown in Fig.~\ref{fig:g_spring}. A force sensor (SingleTact\footnote{https://www.singletact.com/} 10N) has been installed on the index finger tip to measure the force applied by the user. The displacement in the spring's length is estimated by the distance between the index finger tip and the thumb obtained from position sensor reading by using the linear mapping shown in Fig.~\ref{fig:linear_map}. 

\subsubsection{Interaction with virtual springs}
Dexmo gloves provide kinesthetic force feedback on interactions with the virtual objects. The kinesthetic feedback allows to perceive stiffness of an object. DEXMO's API utilizes trigger colliders for the interaction between the virtual hand model and the virtual objects to provide force feedback. Collider components define the shape of a virtual object for the purposes of physical collisions or interactions between the virtual objects. To make interactions computationally efficient, compound colliders are used to roughly approximate the virtual object's mesh using primitive box, cylinder or capsule colliders. This type of interaction is computationally efficient but suffers from a lack of precision and hence is not suitable for the interactions with deformable virtual objects. Another type of colliders are mesh colliders which are precise but computationally expensive. We have used a box (a primitive) collider to define the virtual spring boundary. This collider is used to initiate interaction with collision colliders of the virtual hand. For force feedback, instead of using \textit{Trigger Colliders Hand Model}, we have coded the mesh deformation and the force feedback in response to force applied by the user for interaction with dynamic virtual objects (springs). This code is activated when the virtual spring (box) collider and virtual hand colliders interact. The force feedback with respect to finger position has been calculated by the Hooke's law (\ref{eq:HLaw}). The modifications in the virtual spring's length have been calculated from the index finger position sensor reading using the linear mapping  shown in Fig.~\ref{fig:linear_map}.

\begin{figure*}
    \centering
    \includegraphics[width=7in]{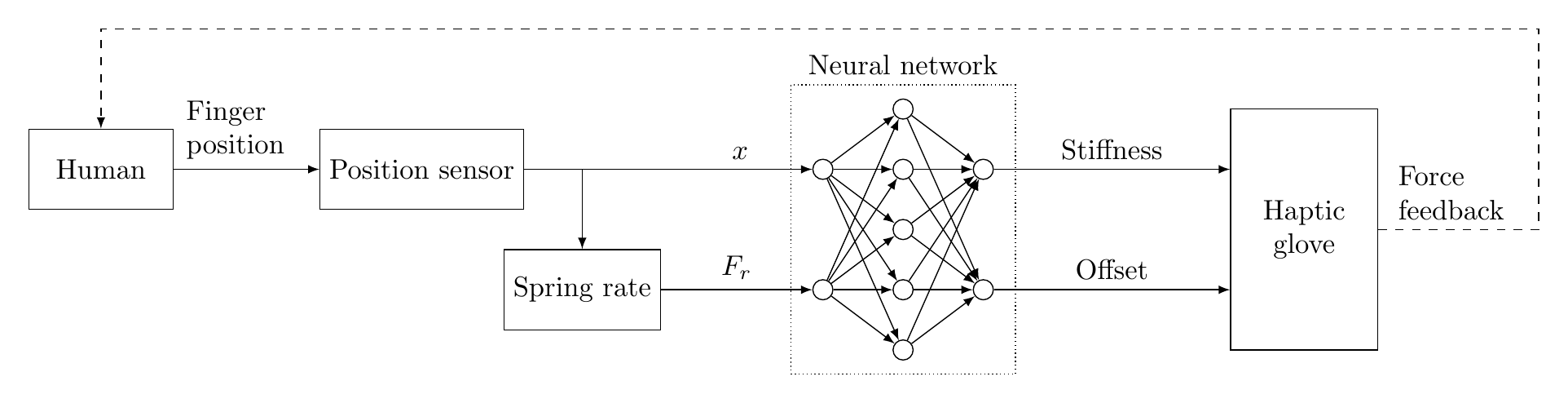}
    \caption{Open loop implementation of neural network}
    \label{fig:NN_open}
\end{figure*}

The force feedback in response to interactions with the virtual spring is controlled directly by the two input parameters (stiffness and position offset) provided by the DEXMO's API. Due to Dexmo’s proprietary force feedback control loop and non-linear motor dynamics, we have used neural network to estimate the relation between the two input parameters and the force applied at index finger tip through glove's motor at MCP joint. Specifically, the neural network provides a mapping between the required force at the finger position (representing displacement in the length of the spring) and motor's internal parameters (stiffness, offset) that control force feedback. The resulting implementation is illustrated by Fig.~\ref{fig:NN_open}. 

The selected neural network is composed of two inputs (finger position and force), one single hidden layer with five neurons, and two outputs (stiffness and offset). This neural network has been trained using the \texttt{Neural Net Fitting tool} (nftool) implemented in \textsc{MATLAB}. The data used to train the neural network have been generated by recording the force values (using force sensor) and position values (using position sensor) for different values of the input parameters (stiffness and offset). Specifically, for different values of the parameters, the user moves the index finger. The resulting force sensors values are saved for different position values. 

\begin{figure*}
    \centering
    \includegraphics[width=5in]{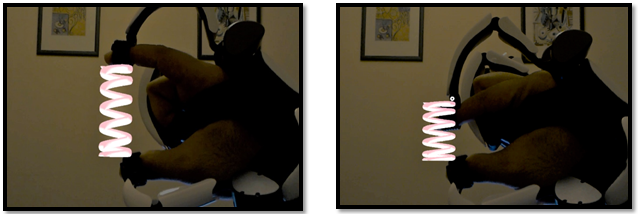}
    \caption{Compressing a virtual spring}
    \label{fig:v_spring}
\end{figure*}

After completion of the training process, the weights and biases of the trained neural network have been directly programmed into the interface. As shown in Fig.~\ref{fig:NN_open}, the position sensor value (displacement $x$) and the desired force ($F=kx$) are used as inputs and the trained neural network sets the values of offset and stiffness to deliver the desired force feedback at the index finger tip. The interaction with the virtual spring is illustrated by Fig.~\ref{fig:v_spring} which shows the user compressing a virtual spring of $k=0.1~\mathrm{N/mm}$ and perceiving a force feedback accordingly.

\section{Results}\label{sec:Results}

This section presents the results obtained with the implementations discussed in the previous sections.

\subsection{Generation of the customized 3D model of an orthopedic cast}

\begin{figure}
    \centering
    \includegraphics[width=3.5in,height=1.5in]{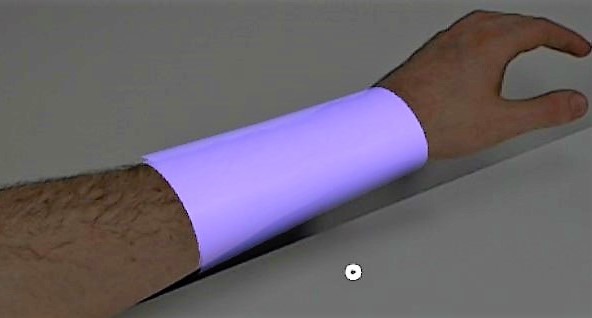}
    \caption{Generated 3D model of the cast}
    \label{fig:vcast}
\end{figure}

\begin{figure}
    \centering
    \includegraphics[width=3.5in]{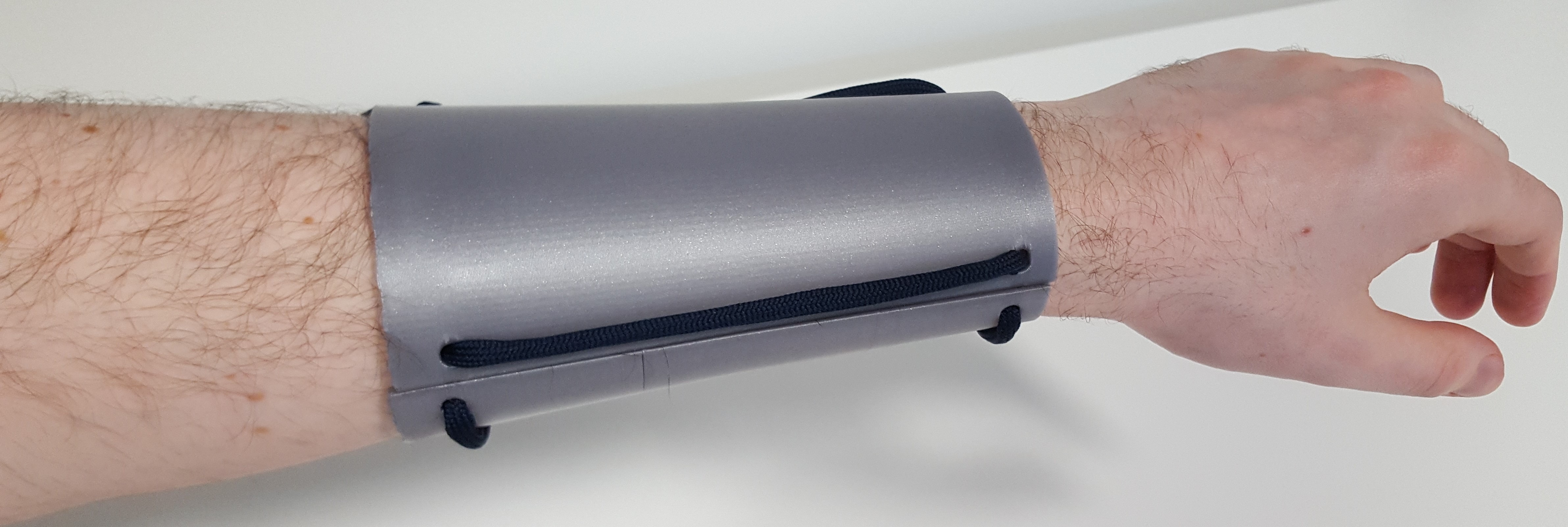}
    \caption{3D printed cast}
    \label{fig:cast}
\end{figure}

The generated 3D model of the cast is shown in Fig.~\ref{fig:vcast}. To make the cast wearable, the generated 3D model is printed in two halves. Many assembly methods, such as snap-fit \cite{klahn2016design}, can be introduced in the generated halves of the 3D model. We have adopted here a simple approach consisting in the introduction of holes (using blender) in both halves. This way, the two halves can be fitted together on the human arm using a thread as shown in Fig.~\ref{fig:cast}.

As the Dexmo glove is under-instrumented, that is, the secondary bar rotations are unknown. This makes the pose estimation of the hand uncertain, which limits the usage of the data from position sensors of all digits to take measurements for scanning. Therefore, we have used only index finger's position sensor with constant pose to take measurements. Future works will include devising a kinematic hand model for the glove that enables the user to utilise all the digits to scan a physical object. Future works will also include the generation of 3D models of the fractured bone using x-Ray scans and augmenting the display of the 3D model onto the patient arm so that the design of the cast can also be customized according to the fractured bone.

\subsection{Elasticity estimation of real spring}

\begin{figure}
    \centering
    \includegraphics[width=3in]{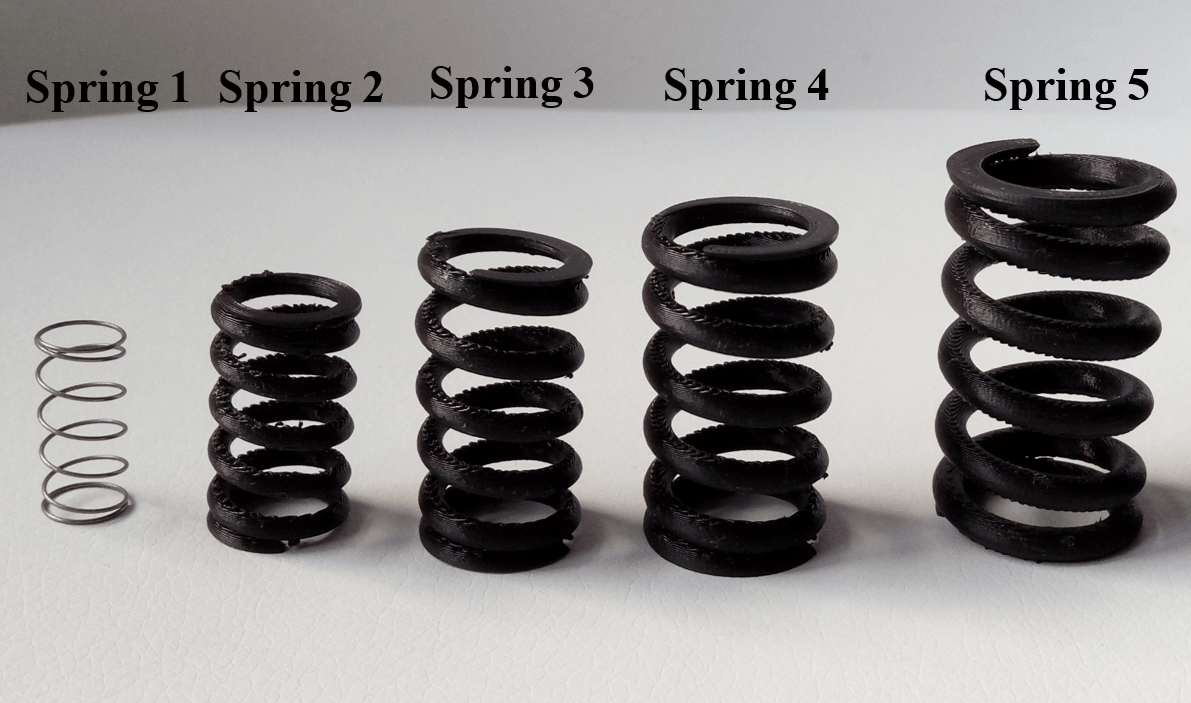}
    \caption{Springs used for validation purposes}
    \label{fig:springs}
\end{figure}

\begin{figure*}
    \centering
    \includegraphics[width=7in]{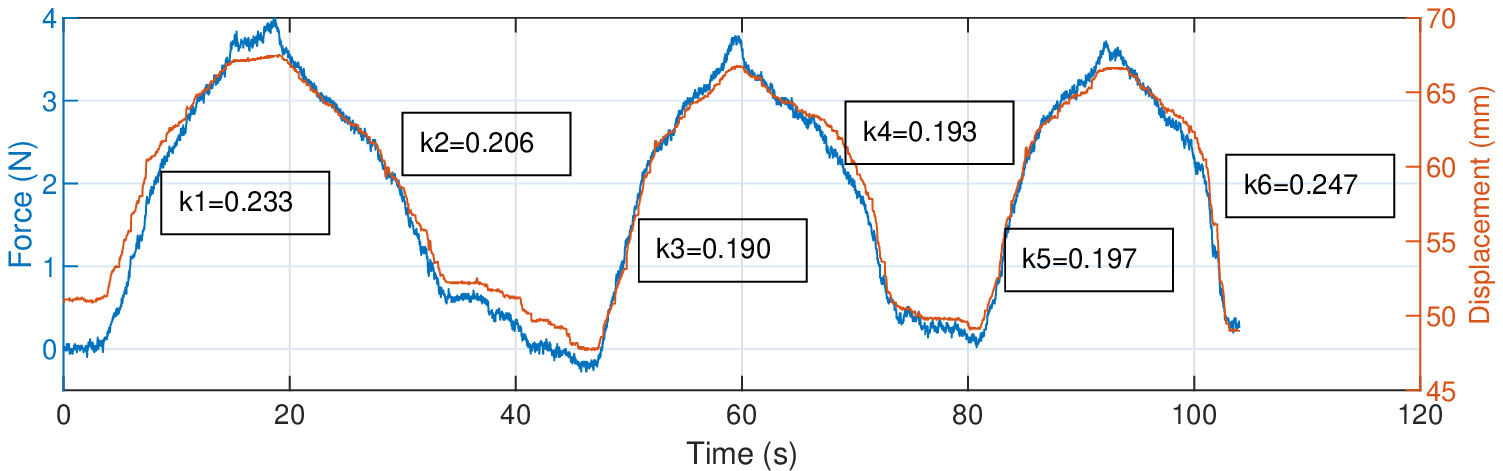}
    \caption{Force-Displacement graph of Spring 1}
    \label{fig:spring1a}
\end{figure*}

\begin{figure}
    \centering
    \includegraphics[width=3.5in]{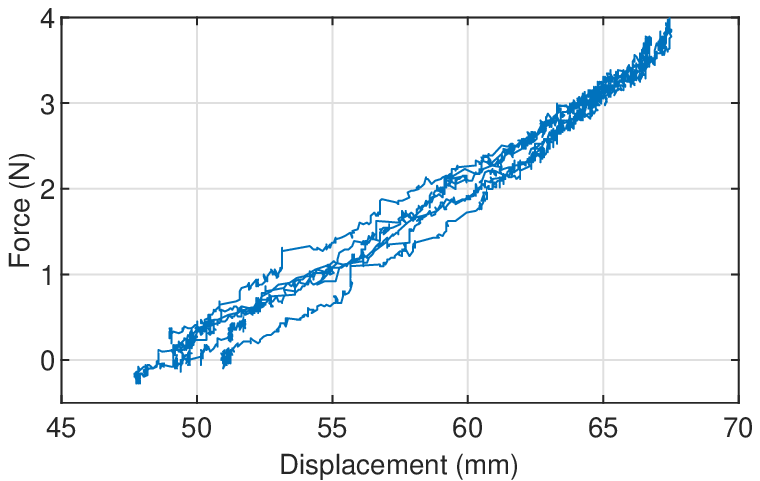}
    \caption{Force vs Displacement graph of Spring 1}
    \label{fig:spring1b}
\end{figure}

The spring rates of the five springs shown in Fig.~\ref{fig:springs} have been estimated using I-nteract. To estimate the spring rate, the spring is compressed multiple times by the user while wearing the glove. The applied force and the resulting displacement in the spring's length are recorded by the force sensor and the position sensor, respectively. Figure~\ref{fig:spring1a} shows the force and displacement data used to estimate the spring rate for Spring 1. In Fig.~\ref{fig:spring1a}, an increase (resp. decrease) of the displacement value indicates a compression (resp. release) cycle of the spring. Figure~\ref{fig:spring1b} shows a quasi linear relationship between the measured force and the displacement data. This quasi linear relationship has been approximated by a line equation using a linear regression. The slope of the line gives us the estimated value of the spring rate. The estimated spring rates of all the springs using I-nteract have been compared with a materials testing machine (Lloyd LR30K\footnote{https://atrya.com.mx/pdf/LR30K.pdf}). The experiment results are given in Tab.~\ref{tab:rates}, assessing the validity of our approach.   

\begin{table}[]
\caption{Estimated spring rates in $\mathrm{N/mm}$}
\label{tab:rates}
\centering
\begin{tabular}{|l|c|c|}
\hline
\multicolumn{1}{|c|}{} & \textbf{\begin{tabular}[c]{@{}c@{}} I-nteract \end{tabular}} & \textbf{\begin{tabular}[c]{@{}c@{}} LR30K\end{tabular}} \\ \hline
Spring 1                              & 0.246                                                                           & 0.243                                                                      \\ \hline
Spring 2                              & 0.42                                                                            & 0.49                                                                       \\ \hline
Spring 3                              & 0.57                                                                            & 0.6                                                                        \\ \hline
Spring 4                              & 0.615                                                                           & 0.643                                                                      \\ \hline
Spring 5                              & 0.677                                                                           & 0.75                                                                       \\ \hline
\end{tabular}
\end{table}

The material of Spring 1 is steel whereas the rest of the springs are 3D printed from PLA filament. This shows that I-nteract can be used to estimate the spring rates of springs with different materials. Therefore, the type of material of the spring can be estimated using: 
\begin{equation*}
    k=\frac{Gd^4}{8nD^3}
\end{equation*}
where $k$ is the spring constant, $G$ is the shear modulus of the material, $d$ is the wire diameter, $D$ is the mean diameter, and $n$ is the number of active coils. The future works include material and elasticity estimation of deformable physical objects.  

\subsection{Simulation of virtual spring}

\begin{figure*}
    \centering
    \includegraphics[width=7in]{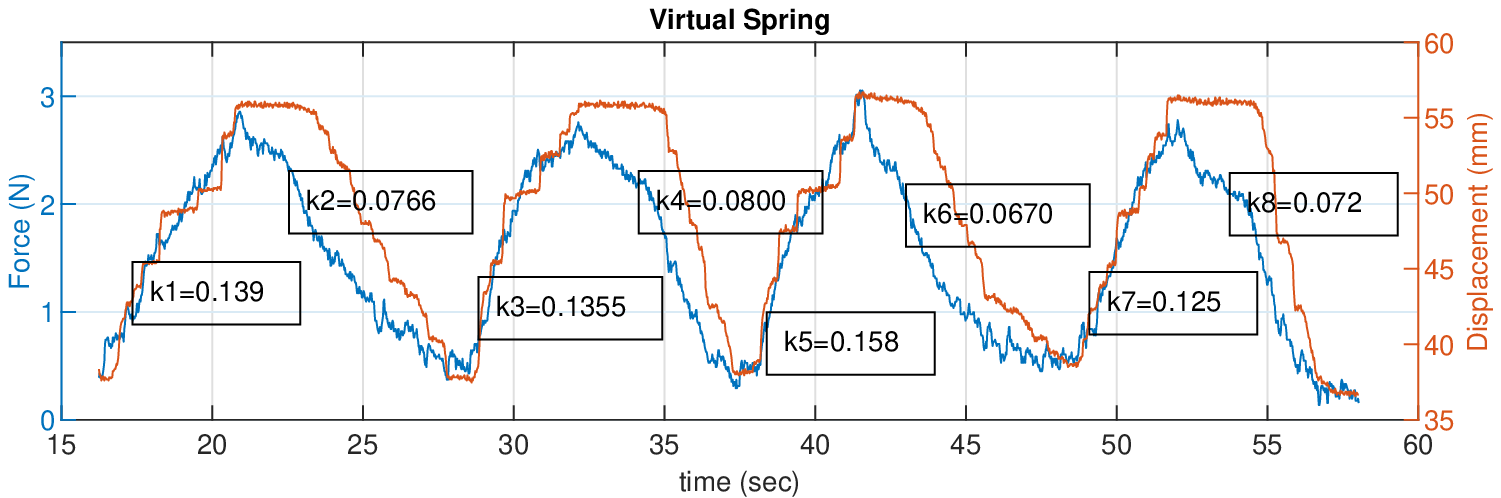}
    \caption{Force-Displacement graph of virtual spring}
    \label{fig:vspring1}
\end{figure*}

\begin{figure}
    \centering
    \includegraphics[width=3.5in]{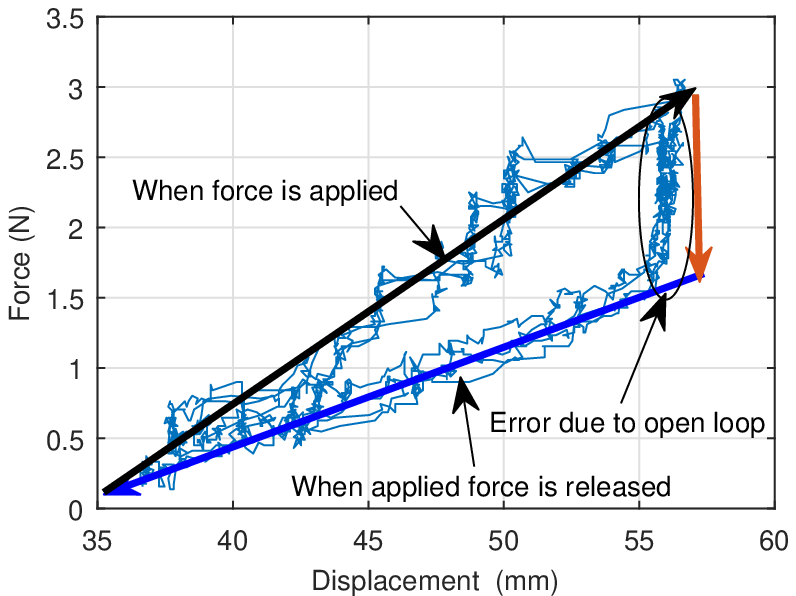}
    \caption{Force vs Displacement graph of virtual spring}
    \label{fig:vspring2}
\end{figure}

Figure~\ref{fig:vspring1} shows the force and displacement data recorded when interacting with virtual spring with a simulated spring rate of $k=0.1~\mathrm{N/mm}$. The discrepancy in the spring rate estimated values between the rising edges (when virtual spring is compressed) and the falling edges (when the applied force is removed) is due to open loop implementation of the neural network. That is the force sensor readings are not fed back to the neural network. The interaction cycle shown in Fig.~\ref{fig:vspring2} consists of the following three steps.
\begin{enumerate}
    \item When the user applies force, the virtual spring's displacement increases quasi-linearly with the force applied.
    \item When the user just starts releasing the applied force, there is sudden drop of almost 1N. This is because, due to open loop implementation the force feedback is not regulated in response to user's applied force.
    \item When the user releases the applied force, the displacement decrease quasi-linearly with the amount of force released but with different rate (slope).
\end{enumerate}
Therefore, Fig.~\ref{fig:vspring2} shows two quasi-linear lines one with slope, $k = 0.12~\mathrm{N/mm}$ during step 1 when the virtual spring is compressed and the other with slope, $k = 0.07~\mathrm{N/mm}$ during step 3 when the applied force is released.

\begin{figure*}
    \centering
    \includegraphics[width=7in]{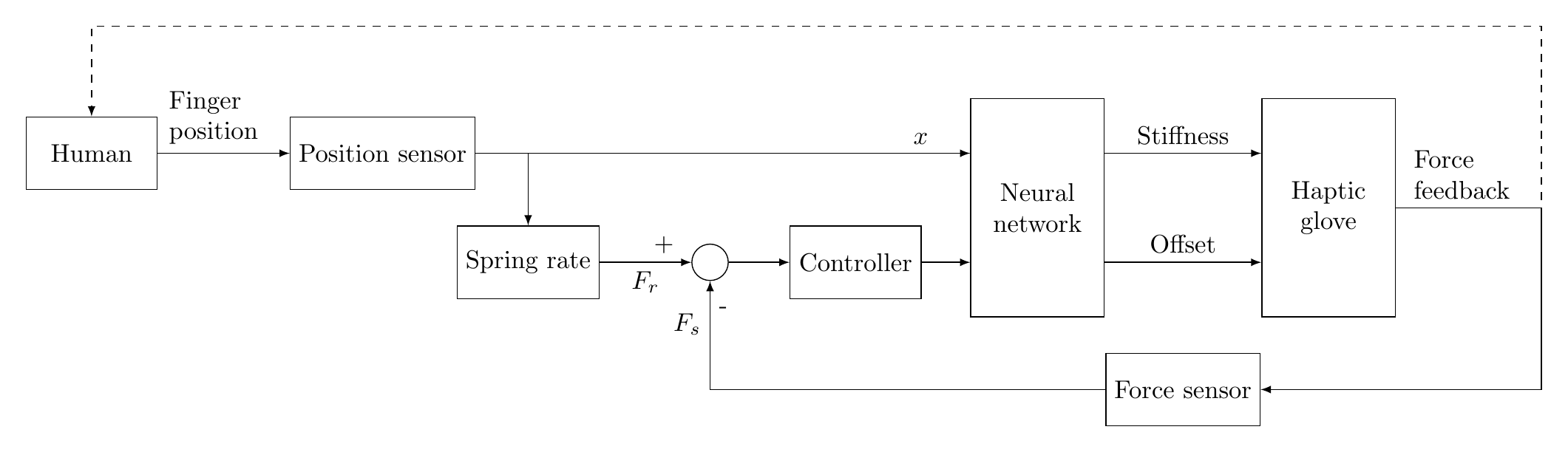}
    \caption{Closed-loop implementation}
    \label{CLNN}
\end{figure*}

The future works include closed-loop implementation of the neural network as shown in Fig \ref{CLNN}. Since mass-spring models are widely used for the implementation of deformable objects simulations~\cite{desbrun1999interactive}, we intend to extend our spring implementation to enable interaction with complex deformable virtual objects using I-nteract.

\section{Conclusion}\label{sec:Conclusion}

In this paper, we presented I-nteract, a cyber-physical system that lies at the frontiers of mixed reality, artificial intelligence, human-machine interaction, robotics, dynamics and control. I-nteract is an interface for real-time interaction with both physical and virtual artifacts in a mixed reality environment to design customized products for personal fabrication. I-nteract enables kinematic and dynamic interaction with physical as well as virtual objects for AM. The kinematic physical interaction efficacy of I-nteract is demonstrated by designing and 3D printing a customized orthopedic cast for a human arm. The dynamic physical interaction is demonstrated by estimating spring constants of real springs of different materials whereas dynamic virtual interaction is demonstrated by simulating a virtual spring. We intend to improve our system by introducing additional features. The objective is to develop an intuitive and user-friendly interface that not only enables 3D models generation and modification in shape and size based on user's inputs through voice commands and hands for personal fabrication, but also provides effective means of monitoring the AM process in a mixed reality environment.

\bibliographystyle{IEEEtran}
\bibliography{IEEEabrv,Library_human_machine_interaction}

\end{document}